\documentclass[sigconf]{acmart}
\copyrightyear{2023}
\acmYear{2023}
\setcopyright{acmlicensed}
\acmConference[CIKM '23] {Proceedings of the 32nd ACM International Conference on Information and Knowledge Management}{October 21--25, 2023}{Birmingham, United Kingdom.}
\acmBooktitle{Proceedings of the 32nd ACM International Conference on Information and Knowledge Management (CIKM '23), October 21--25, 2023, Birmingham, United Kingdom}
\acmPrice{15.00}
\acmISBN{979-8-4007-0124-5/23/10}
\acmDOI{10.1145/3583780.3614866}
\usepackage{multirow}
\usepackage{caption}
\usepackage{subcaption}
\usepackage{graphicx}
\usepackage{float} 

\AtBeginDocument{%
  \providecommand\BibTeX{{%
    \normalfont B\kern-0.5em{\scshape i\kern-0.25em b}\kern-0.8em\TeX}}}

\acmSubmissionID{1048}

\newcommand{\hide}[1]{}
\begin{document}

\title{MERIT: A Merchant Incentive Ranking Model for Hotel Search \& Ranking}
\author{Shigang Quan}
\affiliation{%
  \institution{Shanghai Jiao Tong University}
  \city{Shanghai}
  \country{China}
}
\email{quan123@sjtu.edu.cn}

\author{Hailong	Tan}
\affiliation{%
  \institution{Alibaba Group}
  \city{Hangzhou}
  \state{Zhejiang}
  \country{China}
}
\email{hailong.thl@alibaba-inc.com	}

\author{Shui Liu}

\affiliation{%
  \institution{Alibaba Group}
  \city{Hangzhou}
  \state{Zhejiang}
  \country{China}
}
\email{shui.lius@alibaba-inc.com}

\author{Zhenzhe	Zheng}

\affiliation{%
  \institution{Shanghai Jiao Tong University}
  \city{Shanghai}
  \country{China}
}
\email{zhengzhenzhe@sjtu.edu.cn}

\author{Ruihao Zhu}
\affiliation{%
  \institution{Cornell University}
  \city{Ithaca}
  \country{USA}
}
\email{ruihao.zhu@cornell.edu}

\author{Liangyue Li}
\affiliation{%
  \institution{Alibaba Group}
  \city{Hangzhou}
  \state{Zhejiang}
  \country{China}
}
\email{liliangyue.lly@alibaba-inc.com}

\author{Quan Lu}
\affiliation{%
  \institution{Alibaba Group}
  \city{Hangzhou}
  \state{Zhejiang}
  \country{China}
}
\email{luquan.lq@alibaba-inc.com}

\author{Fan	Wu}
\affiliation{%
  \institution{Shanghai Jiao Tong University}
  \city{Shanghai}
  \country{China}
}
\email{fwu@cs.sjtu.edu.cn}






\renewcommand{\shortauthors}{Quan, et al.}

\begin{abstract}
Online Travel Platforms (OTPs) have been working on improving their hotel Search \& Ranking (S\&R) systems that facilitate efficient matching between consumers and hotels. Existing OTPs focus almost exclusively on improving platform revenue. In this work, we take a first step in incorporating hotel merchants’ objectives into the design of hotel S\&R systems to achieve an incentive loop: the OTP tilts impressions and better-ranked positions to merchants with high quality, and in return, the merchants provide better service to consumers. Three critical design challenges need to be resolved to achieve this incentive loop: Matthew Effect in the consumer feedback-loop, unclear relation between hotel quality and performance, and conflicts between short-term and long-term revenue.

To address these challenges, we propose \textbf{MERIT}, a \textbf{MER}chant \textbf{I}nce\textbf{T}ive ranking model, which can simultaneously take the interests of merchants and consumers into account. We define a new Merchant Competitiveness Index (MCI) to represent hotel merchant quality and propose a new Merchant Tower to model the relation between MCI and ranking scores. Also, we design a monotonic structure for Merchant Tower to provide a clear relation between hotel quality and performance. Finally, we propose a Multi-objective Stratified Pairwise Loss, which can mitigate the conflicts between OTP's short-term and long-term revenue. To demonstrate the effectiveness of MERIT, we compare our method with several state-of-the-art benchmarks. The offline experiment results indicate that MERIT outperforms these methods in optimizing the demands of consumers and merchants. Furthermore, we conduct an online A/B test and obtain an improvement of 3.02\% for the MCI score. Based on these results, we have deployed MERIT online on Fliggy, one of the most popular OTPs in China, to serve tens of millions of consumers and hundreds of thousands of hotel merchants.
\end{abstract}



\maketitle
\section{Introduction}

\begin{figure}[t]
  \centering
  \includegraphics[width=\linewidth]{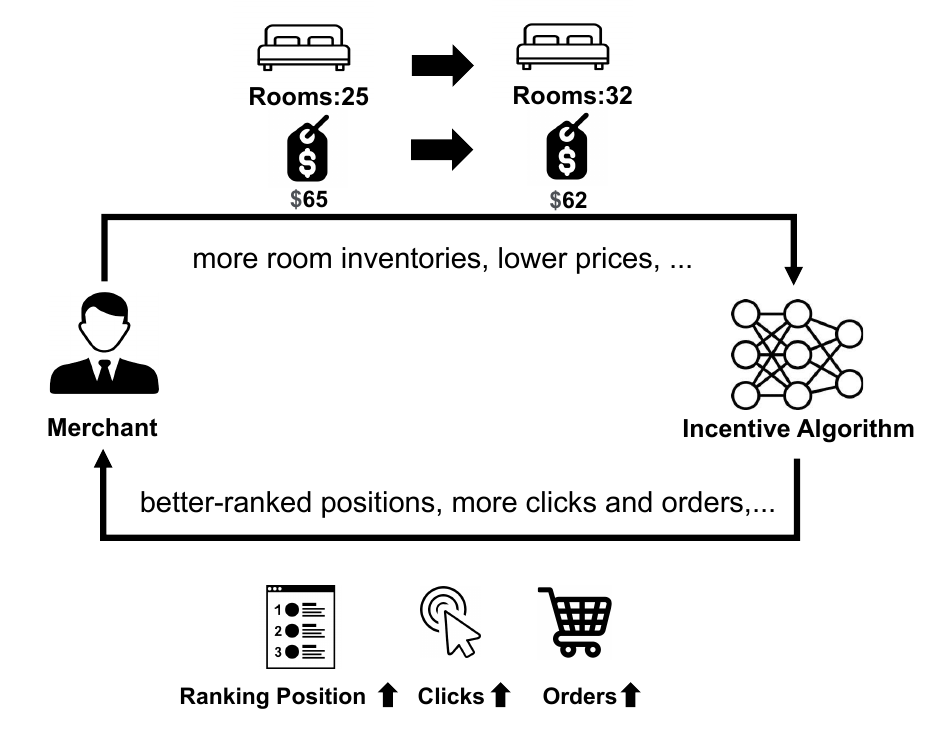}
  \caption{The positive incentive loop for hotel merchants on OTPs.}
   \label{fig:i}
\end{figure}

Nowadays, Online Travel Platforms (OTPs) are applying deep learning \cite{dirn,haldar2019applying,haldar2020improving,g2net,wu2022cheaper} in their hotel Search \& Ranking (S\&R) systems, to accurately provide consumers with their interested hotel-related products. Similar to product recommendation in online e-commerce platform \cite{din,dien}, hotel S\&R systems also need to capture consumers' diverse interests in different hotels to optimize consumer experiences and platform revenue. Different from conventional e-commerce platforms, however, OTPs also have to take the features and objectives of hotel merchants into account. Products sold on OTPs are highly standard room-type items coming from the same hotel merchants, and all OTPs share the same product inventory. As a result, the room inventories and corresponding prices negotiated and acquired from hotel merchants have become the core competitiveness of OTPs in markets. 

To ensure the healthy growth of the OTP and form a good collaborative relation with hotel merchants, it is important to achieve the incentive loop for hotel merchants as illustrated in Figure \ref{fig:i}: the platform tilts impressions and better-ranked positions to the merchants with high quality, and in return, the merchants provide more room inventories and lower prices on the platform, attracting more consumers to the platform. 

However, achieving this incentive loop is not trivial. Though several works \cite{haldar2020improving,wu2022cheaper,zhang2022price} started to consider merchant features, such as the price and location, into hotel S\&R systems, the objective of these works remains to serve consumers and ignore the demands from hotel merchants. Some works \cite{b2006incentive,b2009incentive} consider designing positive incentive mechanisms for consumers. In this work, we take a first step in incentivizing hotel merchants' objectives into the design of hotel S\&R systems. 
Several critical, yet largely overlooked, design challenges during the interaction between consumers and merchants through OTPs are summarized as follows:

$\bullet$ \textbf{Matthew Effect in consumer feedback-loop.} 
On OTPs, impressions and ranked positions of hotels are based on their historical CTRs (Click-Through Rates) and CVRs (Conversion Rates). The consumer feedback about CTRs and CVRs impacted by the current hotel positions will be used for the next ranking stage as a loop~\cite{nips2015feedbackloop}. The above feedback-loop in current hotel S\&R systems would cause Matthew Effect \cite{chen2020bias,wu2022agde}: more impressions and better-ranked positions are tilted toward those hotels with higher historical CTRs and CVRs, and merchants with better quality would not obtain their desired ranked positions due to the lack of historical data. The OTPs need to break the Matthew Effect and provide flexible knobs for potential high-quality merchants to optimize their obtained impressions and ranked positions. 

$\bullet$ \textbf{Unclear relation between hotel quality and performance.} From hotel merchants' perspectives, they hope to receive  explicit and confirmed performance feedback after improving the quality of their hotels. But existing hotel S\&R systems directly take the features of consumers,  hotel rooms and context, as the inputs of black-box learning models. The quality of hotel rooms would be mixed or even under-weighted in various features during the ranking stage. Thus, the hotel merchants cannot know a clear relation between the hotel quality and the potential performance improvement. This unclear relation of hotel feature and performance decreases the incentives for hotel merchants to invest on the OTPs.
    
$\bullet$ \textbf{Conflicts between short-term and long-term revenue.} A consumer's whole behavior cycle can be summarized as follows: a consumer books a hotel online on OTPs, checks in offline, and finally submits ratings on OTPs. OTPs focus almost exclusively on improving CTRs and CVRs, which can be thought of as short-term revenue. However, the quality of merchants can not be reflected by online booking. Ignoring the quality of merchants will deteriorate the long-term revenue of OTPs. Therefore, from the perspective of OTPs, we should improve long-term revenue while striking to ensure that short-term revenue is nearly unchanged.

By jointly considering these challenges, we propose \textbf{MERIT}, a \textbf{MER}chant \textbf{I}nce\textbf{T}ive ranking model, to explicitly represent the relation between hotel merchants and consumers, and optimize the OTP's revenue via the ranking stage in the hotel S\&R system. To tackle the first challenge, we define a new Merchant Competitiveness Index (MCI) metric to represent the quality of hotel merchants, and design a Merchant Tower to model the relation between the MCI and ranking scores. In addition, it will be combined with the CTR Tower and CVR Tower to improve both merchant performance and consumer experiences. To address the second challenge, we introduce a monotonic neural network into the Merchant Tower. The monotonic neural network has been implemented in deep learning to guarantee the model interpretability  \cite{cole2019avoiding,nguyen2019mononet,milani2016fast,feelders2000prior,levi2022nonparametric}. The monotonicity in the Merchant Tower will provide an explicit ranking rule that can be understood by the hotel merchants, and guide them to further optimize their hotel products for higher ranking scores.
To resolve the final challenge, we define a Multi-objective Stratified Pairwise Loss to mitigate conflicts between the short-term and long-term revenue, which guarantees that we can improve the long-term revenue of OTPs without sacrificing the short-term revenue too much. 
Based on the solution described above, MERIT jointly learns the interests and intentions of merchants and consumers, improving the consumer-hotel matching performance of the hotel S\&R system in the long run. We conduct an extensive offline experiment on a large-scale offline hotel S\&R dataset, and results show that MERIT is effective in improving merchant satisfaction and platform performance. In the online A/B test, the average MCI scores have increased by 3.02\% compared with the baseline model. Furthermore, MERIT has been deployed on Fliggy\footnote{\url{http://www.fliggy.com/}}, one of the most popular OTPs in China, where it serves tens of millions of consumers and hundreds of thousands of hotel merchants. We summarize our main contributions in this work as follows:
\begin{enumerate}
 \item We explicitly  consider the quality of hotel merchants in hotel S\&R systems. We define the MCI metric to evaluate hotel quality, and 
 propose a new MERIT model with a Merchant Tower to calculate the MCI metric and integrate it into the ranking scores.
 \item We introduce monotonicity property for the merchant Tower to provide a clear and positive relation between the hotel features and ranking performance. The offline experiment demonstrates that guaranteeing MCI monotonicity and improving monotonicity sensitivity will increase the incentives for hotel merchants to invest on OTPs.
 \item In order to mitigate the conflicts between short-term and long-term revenue of OTPs, we propose the Multi-objective Stratified Pairwise Loss. The extensive offline experiment indicates that the ranking process can be improved via this loss function as opposed to multi-objective pairwise loss.
 \item We conduct both offline evaluation on a large-scale hotel S\&R dataset and a online A/B test. Evaluation results show that MERIT is effective in improving merchant satisfaction and platform performance. Furthermore, MERIT has been deployed on Fliggy and brought remarkable profit growth.
\end{enumerate}

\section{Related Work}
\subsection{Click-through Rate Prediction}
In the search engine, online advertising, and recommendation system, predicting consumers' CTR has become a key problem and a lot of works have focused on this area. Recently, the deep learning methods \cite{cheng2016wideanddeep,guo2017deepfm,wang2017dcn,PNN,he2017neural,din,dien,dirn} have been introduced into CTR prediction models due to the strong representation of deep learning methods. These CTR prediction models often follow the paradigm of Embedding Layer and MLP. To strengthen the capability of capturing the nonlinear feature interaction, some works propose specific networks to learn high-order cross features. Cheng et al. \cite{cheng2016wideanddeep} use the wide structure to memorize nonlinear features and utilize the deep model to improve the generalization of the feature interaction. Guo et al. \cite{guo2017deepfm} propose the DeepFM to combine the traditional factorization machine method \cite{rendle2010factorization} and DNN. Another line of the CTR prediction model is capturing consumer interests from consumer historical behaviors. Zhou et al. \cite{din} propose DIN to utilize the attention mechanism to capture consumers' diverse interests. DIEN \cite{dien} extends DIN and uses the Gated Recurrent Unit (GRU) \cite{chung2014GRU} layer to model the process of sequential interest evolving.

Our research is different from CTR prediction methods in two aspects: First, the CTR prediction model considers the consumer click behavior as its label while our method considers both consumer sequential behaviors and the merchant quality as our labels. Another distinction is that we incorporate and model merchant interests, which is not taken into account in previous work.
\subsection{Multi-task Learning}
Multi-task learning is to simultaneously utilize a composite model to complete different tasks. The original multi-task learning model often takes the shared-bottom layers \cite{caruana1997hard_share} to model the complex relation between various tasks. But this design of shared-bottom layersis strongly constrained. MoE \cite{jacobs1991moe} proposes the gate layer to balance the conflicts of different tasks. MMoE \cite{ma2018mmoe} extends MoE to add multiple gates. Tang et al. \cite{tang2020ple} propose PLE and utilize task-specific experts and task-shared experts to mitigate the negative transfer problem between tasks. The multi-task learning models in the recommendation system and online advertising are often used to predict consumer CTRs and CVRs. Ma et al. \cite{ma2018entire} propose ESMM to tackle the problems of data sparsity and sample selection bias. Wen et al. \cite{wen2020entire} propose the consumer post-click behavior decomposition method to address the aforementioned problems.

Our method also attempts to optimize multi-tasks, but we do not adhere to the basic paradigm of Shared\&Gated Networks because the objectives of consumers and merchants are heterogeneous and the shared structure is inappropriate for this scenario.

\subsection{Monotonicity in Neural Networks}
Adding prior knowledge to a model can reduce the search space of model parameters. As one of prior information, monotonicity can ensure the neural networks can generalize better and improve the interpretability of a black-box neural network. Current adding monotonicity in neural networks can be classified into two types. The first type is constraining the structure of neural networks. The MIN-MAX Network \cite{daniels2010minmax} utilizes a three-layer network to make the model partial monotonic. You et al. \cite{you2017dln} propose Deep Lattice Networks (DLN) and take the ensembles of lattice and calibrators as the constraints of monotonicity. Though DLN can improve the precision, the complexity of the model also increases. Another type is changing the loss function. Gupta et al. \cite{gupta2019pointwiseloss} propose pointwise monotonic loss to penalize the negative gradients. Liu et al. \cite{liu2020certified} add the monotonicity regularization with the uniform data sampling. 

In contrast to former works \cite{haldar2020improving,wu2022cheaper}, which utilize the monotonicity structure to improve the accuracy, in this paper, we follow the monotonicity structure to provide an explicit ranking rule that can be understood by the merchants, and guide them to further optimize their hotels for higher ranking scores.

\section{Preliminaries}
\begin{figure}[t]
  \centering
  \includegraphics[width=\linewidth]{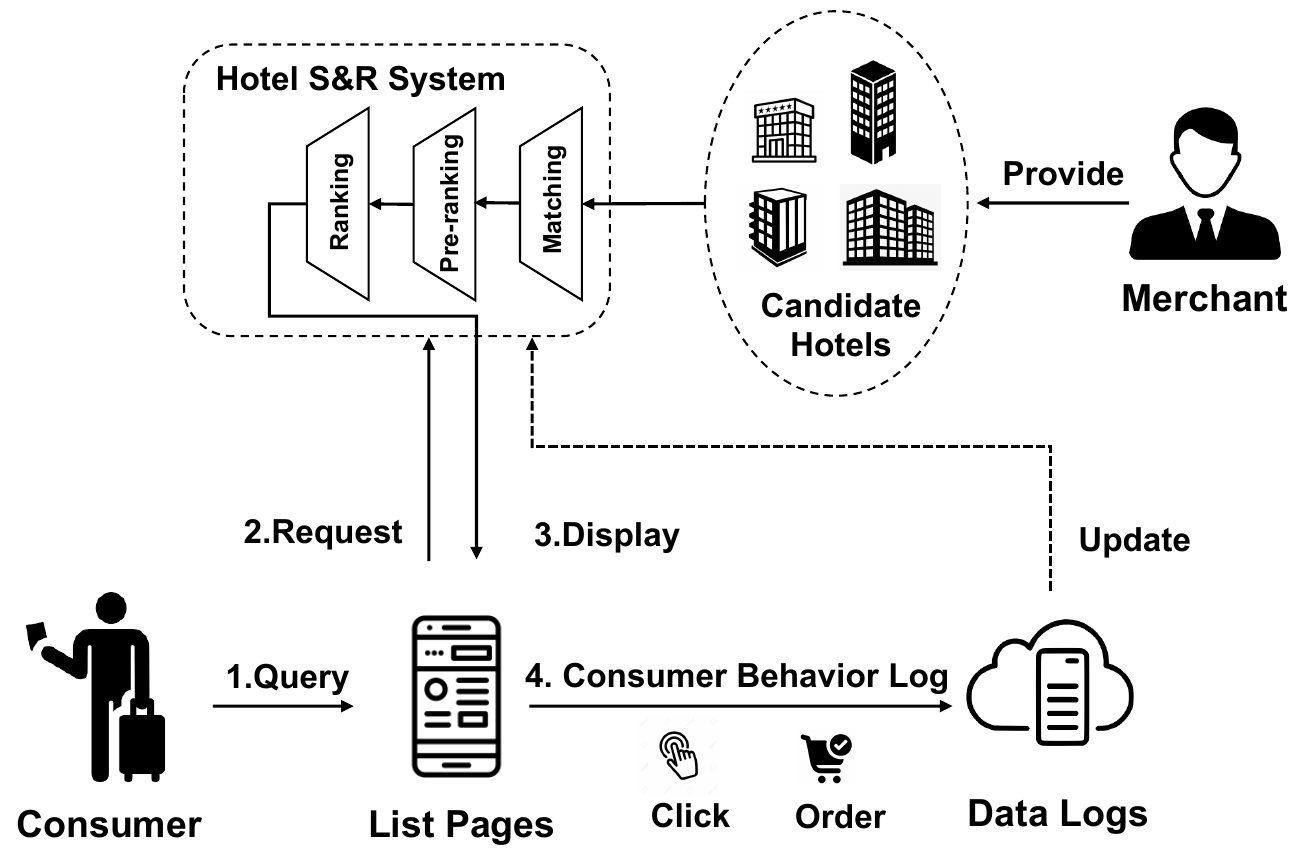}
  \caption{The overview framework of hotel S\&R system.}
   \label{fig:s}
\end{figure}
\subsection{An Overview of Hotel S\&R System }
We briefly introduce the hotel S\&R system illustrated in Figure \ref{fig:s}. In the scenario of the hotel S\&R system, a consumer can issue a query depending on her travel plan. This request will be sent to the hotel S\&R system, which will select the candidate hotels provided by the merchant through different stages based on the amount of data, computational resources, and response latency: matching, pre-ranking, ranking, and so on. After the final ranking stage, the system recommends the top hotels sorted by their ranking scores, and the consumer will click on the corresponding hotel and enter the detail page. The detailed information includes available hotel rooms, prices, and consumer reviews. Orders are then placed by consumers. Each time a consumer interacts sequentially with the hotel S\&R system, her sequential behaviors (click, order, etc.) will be logged in data logs and used to update the parameters of the hotel S\&R system.
\subsection{Merchant Quality Rating Score}
\label{sec:hr}
Merchant quality is a key indicator used to represent the performance of a merchant from different aspects. To better represent merchant quality, we choose factors of merchant quality and define the Merchant Competitiveness Index (MCI) shown in Table \ref{tab:1}.

The Operational Capability (OC) of a merchant is calculated using three indicators: Inventory-to-Sales Ratio, Gross Merchandise Value (GMV), and Historical Conversion Rate. The room sold ratio of hotel is denoted by Inventory-to-Sales Ratio, whilst the overall revenue for hotels is denoted by GMV. The Online Inventory refers to the remaining inventory, while the Hot Selling Room Ratio refers to the remaining hot selling room inventory that the merchant can provide. The Service Refusal Rate is the percentage of hotel orders are rejected when consumers check in offline. The Order Refusal Rate is the percentage of hotel orders that are canceled by merchants. Also, we take the information of detail pages into consideration, such as the Picture Quality and the  Completeness of Hotel Information.


In this paper, the MCI score will be used for both input feature $x_{MCI}$ 
 in Section \ref{sec:3.3} and ranking label $z$ in Section \ref{sec:fr}. 

\begin{table}
  \caption{Factors of Merchant Competitiveness Index.}
  \label{tab:1}
  \resizebox{0.45\textwidth}{!}{
  \begin{tabular}{cc}
    \hline\hline
    Domain of Indicators&Indicators\\
    \hline
    \multirow{3}*{ Operational Capability (OC)
    }& Inventory-to-Sales Ratio\\
    &Gross Merchandise Value (GMV)\\
    &Historical Conversion Rate\\
    \hline
    \multirow{2}*{Room Inventory (RI)}& Online Inventory\\
    &  Hot Selling Room Ratio\\
    \hline
    \multirow{2}*{Quality of Service (QoS)}&Service Refusal Rate\\
    &Order Refusal Rate\\
    \hline
    \multirow{2}*{Basic Information Score (BIS)}&Picture Quality\\
    &Completeness of Hotel Information\\
    \hline\hline
    \end{tabular}
    }
\end{table}

\subsection{Problem Definition}
\label{sec:3.3}
In the scenario of the hotel S\&R system, we assume the dataset to be $\mathcal{D}=\left.\left\{\left(\boldsymbol{x}_{i} \rightarrow y_{i},z_{i}\right)\right\}\right|_{i=1} ^{N}$, and we draw the sample $(\boldsymbol{x} \rightarrow y,z)$ from domain $\mathcal{X}\times\mathcal{Y}\times\mathcal{Z}$, where $\mathcal{X}$ is feature space, $\mathcal{Y}$ and $\mathcal{Z}$ are label spaces, and $N$ is the size of $\mathcal{D}$. $\boldsymbol{x}$ can be defined in the form of $\boldsymbol{x}=(u,q,h_t)$, in which $u$ is the consumer on the OTP, $q$ is the query issued by $u$, and $h_t$ is the target hotel. $y$ represents the click\&order label and $z$ represents the MCI label. $y$ is a three-class label, with $y=0$ indicating not being clicked and ordered, $y=1$ indicating being clicked but not ordered, and $y=2$ indicating being clicked and ordered.  So on the OTP, the objective of the hotel S\&R system is to learn a model $\mathcal{F}$ from the dataset $\mathcal{D}$. $\mathcal{F}$ aims to predict the ranking score $\hat{s}$ for consumer $u$ and target hotel $h_t$ based on the input features $\boldsymbol{x}$:
\begin{equation}
    \hat{s}=p\left(y,z \mid \boldsymbol{x}\right),
\end{equation}
where $\hat{s}$ is the $pCTCVR$ ( $pCTCVR$ will be explained in Section \ref{sec:MERIT}) and is obtained by $\hat{s}^{CTR}\times\hat{s}^{CVR}$.

\section{Methodology}
\begin{figure*}[h]
  \centering
  \includegraphics[width=\linewidth]{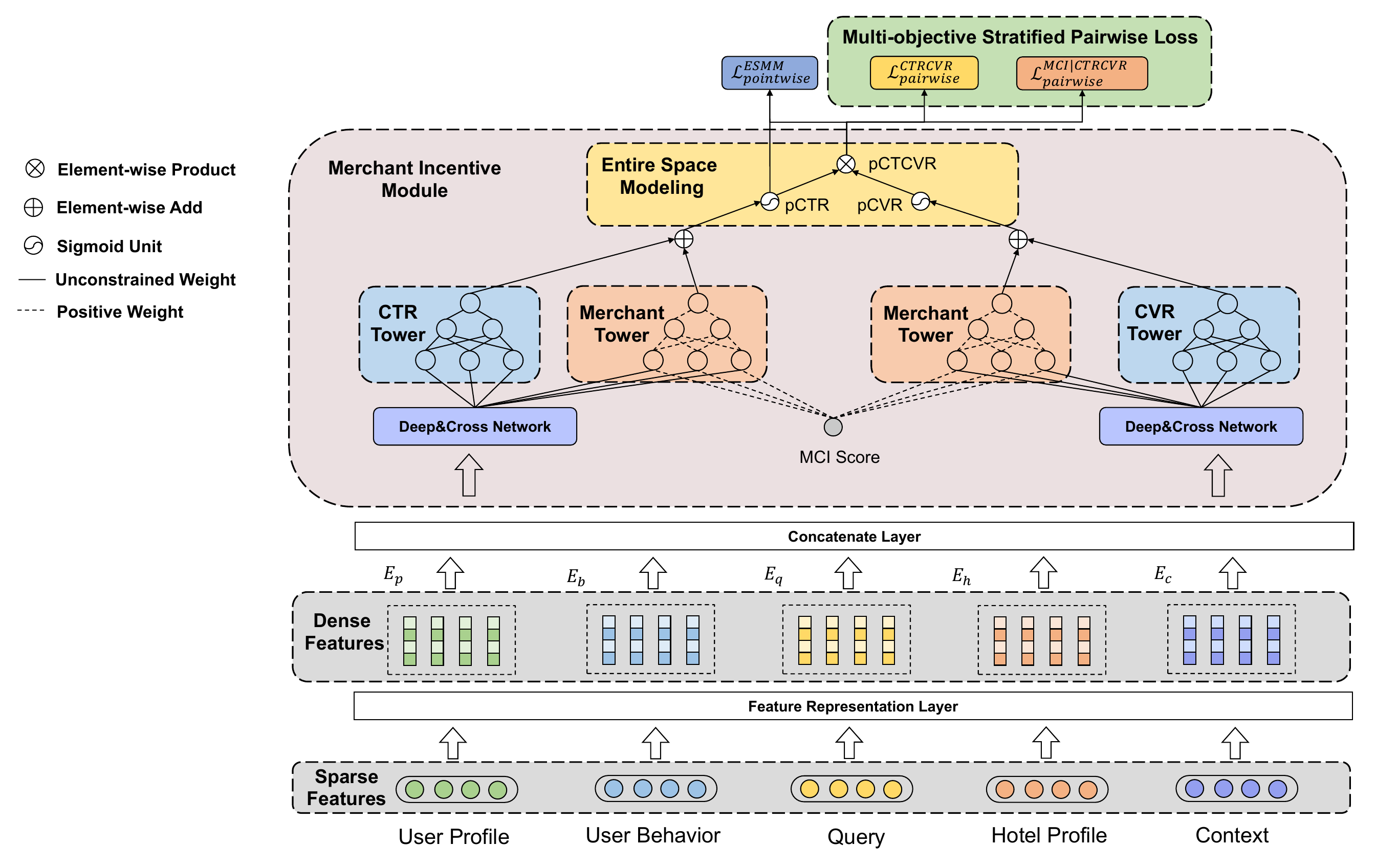}
  \caption{The overview architecture of MERIT, which consists of the Feature Representation Layer, Concatenate Layer, Merchant Incentive Module, and MSPL (Multi-objective Stratified Pairwise Loss). The Merchant Incentive Module consists of Deep\&Cross Network, CTR/CVR Tower, Merchant Tower, and Entire Space Modeling. }
  \label{fig:model}
\end{figure*}

In this section, we give a detailed introduction to our approach as illustrated in Figure \ref{fig:model}. Features and their representation are provided in Section \ref{sec:fr}. Section \ref{sec:MERIT} introduces the framework of the Merchant Incentive Layer and to tackle the conflict of multi-objective pairwise loss, we propose the MSPL in Section \ref{sec:mspl}.
\subsection{Feature Representation}
\label{sec:fr}
We use four types of features: consumer features, query features, hotel features, and MCI features. Consumer features are divided into three categories: consumer basic profile $x^p$ (age, purchase level, etc.), consumer historical preference features $x^b$ (consumer preference on the price and location of a hotel based on her historical interaction behaviors), and consumer context features $x^c$ (time, pid, etc.). The query features $x^q$ include search keywords. The hotel features are hotel item basic profile $x^h$. The features of MCI in Table \ref{tab:1} will be represented as $x^{s}$.

The aforementioned categorical features are encoded as one-hot vectors, while continuous features are discretized and encoded as vectors.  Input vectors are represented by the following symbols: $X^p$, $X^b$, $X^c$, $X^q$, $X^h$, $X^{s}$. To transform these high-dimensional, sparse one-hot vectors into low-dimensional, dense vectors, we utilize embedding layers \cite{mikolov2013word2vec,barkan2016item2vec}. These low-dimensional, dense vectors can be denoted as $(E^{p},E^b,E^{c},E^{q},E^{h})$.
\subsection{Merchant Incentive Module }
\label{sec:MERIT}
To learn the relation between the consumer and the target hotel, we utilize a concatenate layer to concatenate all the embedded feature vectors:
\begin{equation}
    E = [E^p,E^b,E^q,E^h,E^c],
\end{equation}
where $r$ denotes the concatenated embedded vector. Then in order to learn the high-order and low-order interactions of features, we adopt the Deep\&Cross Network \cite{wang2017dcn} into our model as follows:
\begin{equation}
\begin{aligned}
    E^{CTR} = DCN(E),\\
    E^{CVR} = DCN(E),
\end{aligned}
\end{equation}
where $E^{CTR}$ and $E^{CVR}$ denote the representation vectors for CTR and CVR predictions, respectively. For the CTR part, the CTR Tower\footnote{"Tower" is a term used in the field of multi-task learning and typically refers to a neural network designed for individual tasks.} will learn a parameterized mapping function from the representation $E^{CTR}$ to $pCTR$ while the Merchant Tower will learn a parameterized mapping function for $E^{CTR}$ and a monotonic parameterized mapping function for $X^{s}$ as follows:
\begin{equation}
    pCTR = \phi^{B}_{\theta}(E^{CTR},X^{s})+\psi^{CTR}_{\theta}(E^{CTR}),
\end{equation}
where $\phi^{B}_{\theta}(\cdot)$ is a positive-weight feed-forward network 
 \cite{archer1993application} and $\psi^{CTR}_{\theta}(\cdot)$ is a feed-forward network. The Merchant Tower is designed to calibrate the predicting process and strengthen the monotonicity of our model. And CVR prediction structure is similar to the CTR prediction structure as follows:
\begin{equation}
    pCVR =\phi^{B}_{\theta}(E^{CVR},X^{s})+ \psi^{CVR}_{\theta}(E^{CVR}),
\end{equation}
where $\psi^{CVR}_{\theta}(\cdot)$ is a feed-forward network. The relation between the features of MCI and the final ranking score will adjust based on the context, so $E^{CTR}$ and $E^{CVR}$ will be also fed into the Merchant Tower ($\phi^{B}_{\theta}(\cdot)$) to cross $X^{s}$ and other features.

In order to avoid sample Selection Bias (SSB) and Data Sparsity (DS) issues \cite{ma2018entire}, we follow the entire space modeling \cite{ma2018entire}. We calculate $pCTCVR$ (predicted post-view Click-Through rate\&ConVersion Rate) as our final ranking score, and its definition is as follows:
\begin{equation}
     pCTCVR = pCTR \times pCVR.
\end{equation}
\subsection{Multi-objective Stratified Pairwise Loss}
\label{sec:mspl}
Unlike pointwise loss, which calculates the numerical gap of each sample, pairwise loss \cite{rendle2012bpr} measures the range between positive and negative samples. Pairwise loss outperforms pointwise loss in representing the overall ranking performance. Consequently, we incorporated pairwise loss to evaluate the ranking outcomes. However, a conflict may arise when undertaking multi-objective optimization using pairwise loss:

\textbf{Conflict of Multi-objective Pairwise Loss} For the sake of simplicity, we will use two objectives, $Y$ and $Z$, as an example, where $Y$ is the primary objective and $Z$ is the secondary objective. And for each objective, $\mathcal{Y}^{+}$, $\mathcal{Y}^{-}$ are the positive and negative sample sets of $Y$. $\mathcal{Z}^{+}$ and $\mathcal{Z}^{-}$ are the positive and negative sample sets of $Z$. Pairwise loss functions can be defined as follows:
\begin{equation}
\begin{split}
    \mathcal{L}^{Y}_{pairwise}=\sum^{N}_{i=1}\sum^{N}_{j=1}\ell(\hat{s}_{i}-\hat{s}_{j},\mathbb{I}(Y_{i}>Y_{j})),\\
    \mathcal{L}^{Z}_{pairwise}=\sum^{N}_{i=1}\sum^{N}_{j=1}\ell(\hat{s}_{i}-\hat{s}_{j},\mathbb{I}(Z_{i}>Z_{j})), 
\end{split}
\end{equation}
where $\mathbb{I}$ is the indicator function and $\ell (\cdot)$ is the negative log-likelihood loss function. Suppose that if $i \in \mathcal{Y}^{+} \cap \mathcal{Z}^{-}$ and $j \in \mathcal{Y}^{-} \cap \mathcal{Z}^{+}$, then the conflict problem occurs: For objective $Y$, the pairwise loss function $\mathcal{L}^{Y}_{pairwise}$ is encouraged to enlarge the ranking score gap between $i$ and $j$, while for $Z$, the pairwise loss function $\mathcal{L}^{Z}_{pairwise}$ is encouraged to enlarge the ranking score gap between $j$ and $i$. During the training process, inconsistency in gradient directions will deteriorate the effectiveness of multi-objective optimization. In the scenario of the hotel S\&R system, the above conflict corresponds to Challenge 3: the conflict of platform revenue and consumer experience.

\textbf{Multi-objective Stratified Pairwise Loss} In order to tackle the above conflict problem, we define the Multi-objective Stratified Pairwise Loss (MSPL) for the multi-objective ranking. It can be defined as follows:
\begin{equation}
\mathcal{L}^{Z|Y}_{pairwise} = \sum^{N}_{i=1}\sum^{N}_{j=1} \mathbb{I}(Y_{i} \geq Y_{j})\ell(\hat{s}_{i}-\hat{s}_{j},\mathbb{I}(Z_{i}>Z_{j})).
\label{equ:mspl}
\end{equation}
The above pairwise loss function $\mathcal{L}^{Z|Y}_{pairwise}$ can be effective when the second objective $Z$ is consistent with the primary objective $Y$, and, otherwise, the loss function will be masked. And it is noted that samples $i$ and $j$ are symmetrical, so we only consider the greater relation.

In the hotel scenario, $y$ (corresponding $Y$) is the primary objective and $z$ (corresponding $Z$) is the secondary objective, so the loss function can be defined as follows:
\begin{equation}
\begin{aligned}
    \mathcal{L}^{CTRCVR}_{pairwise} &= \sum^{N}_{i=1}\sum^{N}_{j=1} 
    \ell(\hat{s}_{i}-\hat{s}_{j},\mathbb{I}(y_i>y_j)),\\
    \mathcal{L}^{MCI|CTRCVR}_{pairwise}   &=  \sum^{N}_{i=1}\sum^{N}_{j=1} \mathbb{I} (y_i\geq y_j)\ell(\hat{s}_{i}-\hat{s}_{j},\mathbb{I}(z_i>z_j)),
\end{aligned}
\label{equ:mspl}
\end{equation}
 where $\mathcal{L}^{CTRCVR}_{pairwise}$ (corresponding $\mathcal{L}^{Y}_{pairwise}$) a pairwise ranking loss, which aims to enlarge the ranking score gap between $i$ and $j$ when label level $y_i$ is larger than $y_j$. $\mathcal{L}^{MCI|CTRCVR}_{pairwise}$ (corresponding $\mathcal{L}^{Z|Y}_{pairwise}$) is also a pairwise ranking loss for label $z$, but we add an extensive stratified constraint $\mathbb{I} (y_i \geq y_j)$ to ensure the loss is only effective when label level $y_i$ is larger or equal than $y_j$.

We follow the ESMM pointwise loss function \cite{ma2018entire} as the basic loss function:
\begin{equation}
\begin{aligned}
    \mathcal{L}^{ESMM}_{pointwise} &= \sum^{N}_{k=1} \ell(\hat{s}^{CTR}_{k},\mathbb{I}(y_k>0)) +\sum^{N}_{k=1} \ell(\hat{s}_{k},\mathbb{I}(y_k=2)),\\
\end{aligned}
\label{equ:esmm}
\end{equation}
where $\hat{s}^{CTR}_{k}$ denotes the predicted CTR score of sample $k$. And training loss function $\mathcal{L}$ consists of three parts: $\mathcal{L}^{ESMM}_{pointwise}$, $\mathcal{L}^{CTRCVR}_{pairwise}$, and $\mathcal{L}^{MCI|CTRCVR}_{pairwise}$:
\begin{equation}
    \mathcal{L}= \mathcal{L}^{ESMM}_{pointwise} + \lambda_{1} \mathcal{L}^{CTRCVR}_{pairwise}+ \lambda_{2} \mathcal{L}^{MCI|CTRCVR}_{pairwise},
\end{equation}
where $\lambda_{1}$ and $\lambda_{2}$ are hyper-parameters that balance the above three loss functions. We empirically explore their influence in Section \ref{sec:hp}.

\section{Experiments}
To evaluate the effectiveness of our proposed method, we compare it with state-of-the-art methods and report our experimental results and corresponding analysis in this section. In order to choose the appropriate hyper-parameters, we explain the detailed process of choosing $\lambda_1$ and $\lambda_2$. Finally, an online A/B test is conducted in the hotel S\&R system on Fliggy.
\begin{table*}[htb]
    \centering
    \caption{Comparison of multi-task learning models and monotonic networks on the offline hotel S\&R dataset. Results of CTR (Click-Through Rate), CVR (Conversion Rate), and CTCVR (post-view Click-Through\&Conversion Rate) are presented. The best results of all methods are indicated in bold, while the second best results are indicated in underlined. The Gain means the AUC improvement of MERIT+MSPL compared with DNN. }
    \begin{tabular}{ccccccc}
    \hline\hline
    \multirow{2}*{Models}&\multicolumn{3}{c}{AUC}&\multicolumn{3}{c}{GAUC}\\
    \cmidrule(r){2-4} \cmidrule(r){5-7}
    &CTR AUC&CVR AUC&CTCVR AUC&CTR GAUC&CVR GAUC&CTCVR GAUC  \\ \hline
    DNN (Base) & 0.7454&	0.8870&	0.8966&	0.7737&	\underline{0.8106}&	0.8201  \\ 
    Shared Bottom& 0.7453&	0.8767&\textbf{	0.8979}&	0.7727&	0.8014&	\underline{0.8210}\\
    MMoE &\underline{0.7463}&	0.8787&	\underline{0.8975}&	0.7724&	0.8024&	\textbf{0.8216} \\
    CGC &\textbf{0.7464}&	\textbf{0.8885}&	0.8972&	0.7720&	0.7978&	0.8206\\
    MERIT&0.7453&\underline{	0.8882}&	0.8969&	0.7734&	0.8093&	0.8195\\
    MERIT (Point-wise Monotonic Loss)&0.7451&	\textbf{0.8885}&	0.8970&	0.7735&	\textbf{0.8114}&	0.8206\\
    MERIT (MIN-MAX Network)&0.7455&	0.8880&	0.8970&\textbf{	0.7740}&	0.8098&	0.8202\\
    MERIT+MPL&0.7452&	0.8820&	0.8940&	0.7737&	0.8039&	0.8178\\
    MERIT+MSPL&0.7452&	0.8824&	0.8943&	\underline{0.7739}&	0.8049&	0.8188\\
    Gain&-0.0002&-0.0046&-0.0023&+0.0002&-0.0057&-0.0013\\

    \hline\hline
    \end{tabular}
\label{table:auc}
\end{table*}
\hfill
\begin{table*}
    \centering
    \caption{Comparison of multi-task learning models and monotonic networks on the offline hotel S\&R dataset. The MCI ranking results are presented. The best results of all methods are indicated in bold, while the second best results are indicated in underlined. The Gain means the NDCG improvement of MERIT+MSPL compared with DNN and $*$ means p-value $<$ 0.001 in significance tests compared to the best baseline.}
    \begin{tabular}{ccccccc}
    \hline\hline
    \multirow{2}*{Models}&\multicolumn{3}{c}{NDCG}&\multicolumn{3}{c}{wNDCG}\\
    \cmidrule(r){2-4} \cmidrule(r){5-7}
    &NDCG@5&NDCG@10&NDCG@20&wNDCG@5&wNDCG@10&wNDCG@20  \\ \hline
    DNN (Base) &0.8468&	0.8804&	0.9067&	0.7529&	0.7844&	0.8228\\
    Shared Bottom &0.8458&	0.8793&	0.9057&	0.7523&	0.7833&	0.8213\\
    MMoE & 0.8462&	0.8794&	0.9057&	0.7526&	0.7833&	0.8210 \\
    CGC &0.8446&	0.8781&	0.9047&	0.7500&	0.7811&	0.8193\\
    MERIT&0.8525&0.8848&0.9102&0.7615&0.7921&0.8293\\

    MERIT (Point-wise Monotonic Loss)&0.8485 &	0.8818 &	0.9078 &	0.7555&	0.7869&	0.8248\\
    MERIT (MIN-MAX Network)&0.8499&	0.8829&	0.9086&	0.7579&	0.7890&	0.8267\\
    MERIT+MPL&\underline{0.8708}&\underline{	0.9011}&\underline{	0.9235}&\underline{	0.7838}&	\underline{0.8156}&	\underline{0.8516}\\
    
    MERIT+MSPL&$\textbf{0.8718}^{\ast}$&$\textbf{0.9020}^{\ast}$&$\textbf{0.9243}^{\ast}$&$\textbf{0.7850}^{\ast}$&$\textbf{	0.8171}^{\ast}$&$\textbf{0.8531}^{\ast}$ \\
    
    Gain&+0.0250&+0.0216&+0.0176&+0.0321&+0.0327&+0.0303\\

    \hline\hline
    \end{tabular}
\label{table:ndcg}
\end{table*}


\begin{figure*}[htbp]
  \centering
  \includegraphics[width=\linewidth]{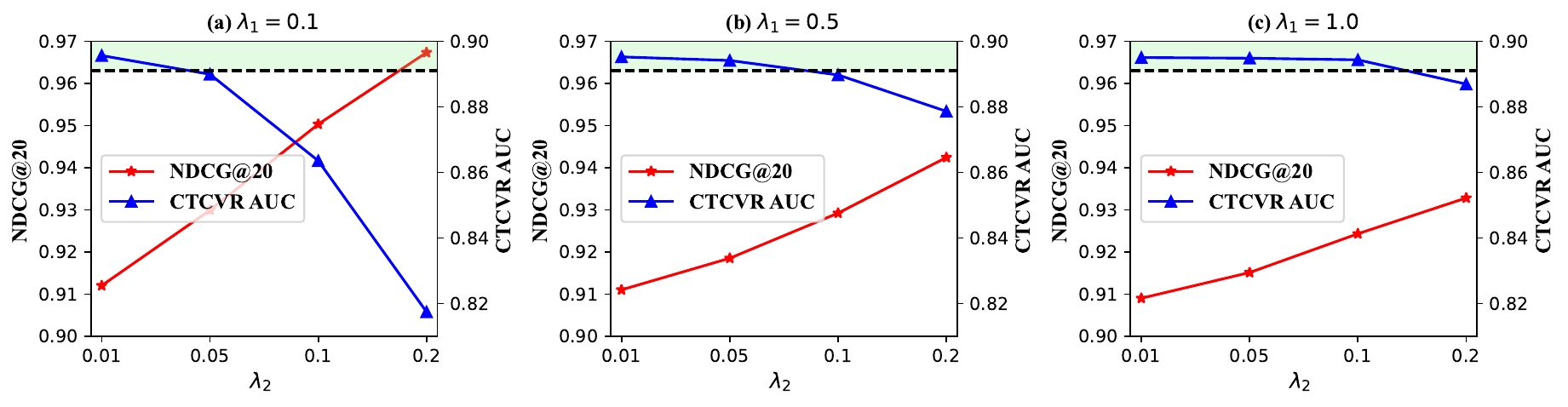}
  \caption{The NDCG@20 of MCI ranking and CTCVR AUC for different hyper-parameters $\lambda_{1}$ and $\lambda_{2}$. The dashed line indicates the lower bound of CTCVR AUC we can tolerate (From the online experiment, we consider that a 0.005 decline of CTCVR AUC is acceptable, so we set the lower bound of CTCVR AUC as 0.892 with the best 0.897.), and we choose $\lambda_1$ and $\lambda_2$ from points in the light green area.}
  \label{fig:hp}
\end{figure*}

\subsection{Dataset and Experimental Settings}
\textbf{Dataset Descriptions}
The offline hotel S\&R dataset is generated based on the consumer logs collected from the hotel S\&R system on Fliggy. As illustrated in Figure \ref{fig:s}, each sample in this dataset is based on the impression of each consumer's search result. We take the clicking and ordering samples as positive samples, while others as negative samples. The features of these samples primarily contain three aspects of information, namely consumers, hotels, and search keywords. The consumer features include basic consumer attributes, long-term and short-term hotel preferences, and the distance between the consumer and the hotel. The hotel features include hotel basic attributes, real-time prices and inventories, historical behavior statistics, and factors of hotel quality. The keyword features include searching scene types, the distance between searching location and candidate hotels, the semantic similarity between search words and candidate hotels, etc. The offline hotel S\&R dataset statistics are presented in Table \ref{tab:sod}. The dataset contains 300 million training samples and 14 million test samples, which are partitioned by time.

\textbf{Experimental Settings}
MERIT is deployed in TensorFlow\footnote{\url{http://tensorflow.org/}}. We employ a grid search strategy to determine optimal hyper-parameters for our models. The ADAM optimizer is used to train all models, with a fixed learning rate of 0.001 and a batch size of 2048. To prevent overfitting, we apply $L_2$ regularization with a weight of 0.00001, as well as batch normalization with a dropout rate of 0.3. The balancing hyper-parameters ${\lambda}_1$ and ${\lambda}_2$ we use are 1.0 and 0.1. The embedding size for consumer and query features is set at 4, while the embedding size for context and item features is set at 8. The CTR and CVR Towers are implemented via a three-layer fully connected network with sizes of 256, 128, and 64. For multi-task learning models, a three-tower MLP is constructed to predict CTR, CVR, and MCI, with the gated network implemented via a softmax layer. MMoE utilizes 8 expert networks, while CGC employs 1 shared expert network and 1 specific expert network. The MIN-MAX network uses 10 groups of 10 linear functions.
\begin{table}[h]
\caption{Statistics of the offline hotel S\&R dataset.}
\begin{tabular}{cccc}
\toprule[0.8pt]
{Category}&{\#User}&{\#Hotel}&{\#Impression}\\
\midrule
{Number}&{5,614,476 }&{775,190 }&{354,319,050 }\\
\midrule
\midrule
{Category}&{\#Click}&{\#Conversion}\\
\midrule
{Number}&{31,326,730 }&{2,317,167 }\\
\bottomrule[0.8pt]
\end{tabular}
\label{tab:sod}
\end{table}

\textbf{Baselines} To evaluate the effectiveness and superiority of our methodology, we compare our method with state-of-the-art multi-task learning models. And we also choose some monotonic networks to compare the ranking results. These baselines are as follows:
\begin{itemize}
    \item \textbf{DNN} \cite{gardner1998mlp}:
     We only use a Multi-Layer Perceptron to construct the CTR Tower and CVR Tower, and input features are the same as MERIT.
    \item \textbf{Shared Bottom} \cite{caruana1997hard_share}:
    The Shared Bottom model uses the unified bottom layer and different towers for all the tasks. It aims to utilize a shared bottom layer to learn correlations between different tasks.
    \item \textbf{MMoE} \cite{ma2018mmoe}:
    The MMoE method with multiple gate networks is designed to control the expert networks for different tasks and relieve the negative transfer problem. 
    \item \textbf{CGC} \cite{ple}:
    Customized Gate Control (CGC) with an Expert-Bottom layer intends to mitigate the seesaw phenomenon \cite{ple} via task-specific experts and task-shared experts. For a fair comparison, we do not take the progressive layered designing.
    \item \textbf{MERIT (MIN-MAX Network)} \cite{daniels2010minmax}: It follows the MERIT structure and replaces the Merchant Tower with the MIN-MAX Network. The MIN-MAX Network is a three-layer structure with max-pooling and min-pooling.
    \item \textbf{MERIT (Point-wise Monotonic Loss)} \cite{gupta2019pointwiseloss}: It follows the MERIT structure and replaces the Merchant Tower with a Multi-Layer Perceptron and Point-wise Monotonic Loss. Point-wise Monotonic Loss is the loss function penalizing the negative gradients.
    \item \textbf{MERIT+MPL}: It follows the MERIT structure and we add Multi-objective Pairwise Loss.
    \item \textbf{MERIT+MSPL}: It follows the MERIT structure and we add Multi-objective Stratified Pairwise Loss as defined in Equation (\ref{equ:mspl}).
\end{itemize}

\textbf{Evaluation Metrics}
Our approach is to strike a balance between consumer and merchant engagement. Specifically, our model focuses on binary classification of whether a hotel has been clicked and ordered by consumers, while also ranking hotels according to MCI for hotels. To evaluate our method against baselines, we employ two types of metrics.:
\begin{itemize}
    \item \textbf{AUC}: \textit{Area Under Curve} (AUC) is a widely used metric to measure the ranking result for the whole model. It indicates the probability that positive samples rank higher than negative samples.
    \item \textbf{GAUC}: \textit{Group Area Under Curve} (GAUC) partitions test samples into groups via the consumer id, and AUC is calculated by each group. It is defined as follows:
\begin{equation}
    \operatorname{GAUC}=\frac{\sum_{u} w_{u} \times \operatorname{AUC}_{u}}{\sum_{u} w_{u}},
\end{equation}
where $w_u$ is the sample size of consumer $u$.
\end{itemize}
\begin{itemize}
    \item \textbf{NDCG@K}: \textit{Normalized Discounted Cumulative Gain} (NDCG) indicates the ratio between current ranking performance and the ideal ranking performance. It considers the ranking position in terms of relation score. In this experiment, the ranking performance totally via the MCI label will be the ideal ranking performance. We choose Top-K test samples as the evaluated group.
    \item \textbf{wNDCG@K}: \textit{weighted Normalized Discounted Cumulative Gain} (wNDCG) partitions the test samples into each group via the session id, and for each session, we calculate the NDCG. Its mathematical formulation is as follows:
    \begin{equation}
        \operatorname{wNDCG@K} =\frac{\sum_{s} w_{s} \times \operatorname{NDCG@K}_{s} }{\sum_{s} w_{s}},
    \end{equation}
    where $w_{s}$ is the length of session $s$.
\end{itemize}

\subsection{Offline Comparison Results}
In this subsection, we compare our proposed model with several multi-task learning models and monotonic networks on the test set of the offline hotel S\&R dataset. The AUC and GAUC results of predicting consumer feedback are reported in Table \ref{table:auc}, while NDCG and wNDCG results of ranking MCI label are illustrated in Table \ref{table:ndcg}. Based on our analysis of predicting consumer implicit feedback, we made the following observations:
\begin{itemize}
    \item Compared with the DNN model, multi-task learning models can improve part of the predicted scores. Specifically, the CGC model achieves the best performance compared with the DNN model and other multi-task learning models, because it follows the different designs of expert networks.
    \item The MERIT model can get nearly equal results compared with the base model. Especially, the CVR AUC improves by 0.0012, 0.0015, and 0.0010 for different monotonic models compared with the base model, which indicates the monotonicity for $X^{s}$ can better predict consumers' purchase behaviors. But the MCI is weakly related to the click behaviors, so the improvement is not noticeable.
    \item When we add MSPL into the MERIT model, some metrics will decrease. But this decrease is modest and tolerable. The objective of our method is to optimize the merchant demand, so the minor decline in consumer demand is unavoidable and reasonable. Also, we can recognize that the decline is primarily due to CVR ranking results.
\end{itemize}

In order to obtain the ranking results of hotel quality, we compare the listed hotel MCI level for different methods. And we summarize our assessments of hotel quality ranking and draw the following findings:
\begin{itemize}
    \item When compared to the DNN model, multi-task learning models do not take hotel quality into account, hence the NDCG shows no discernible improvement. In addition, the relation between consumers and merchants may be negative and insignificant, so the NDCG of multi-task learning models falls dramatically. Specifically, for CGC, which has the best performance in multi-task learning models, the NDCG@5 declines by 0.0022.
    \item The monotone structure has the potential to improve hotel ranking results. In comparison to the base model, three monotonic structures (MERIT, MIN-MAX Net, and Point-wise Monotonic Loss) enhance NDCG@5 by 0.0057, 0.0031, and 0.0017. Because of the extensive batch norm component, our proposed MERIT approach has the best performance.
    \item The Multi-objective Pairwise Loss is capable of strengthening the training objective on the MCI. Specifically, MERIT+ MPL and MERIT+MSPL improve on NDCG@5 by 0.0240 and 0.0250, respectively. The MSPL can mitigate the conflict problem of multi-objective optimization and hence outperform MPL in the ranking of MCI.
\end{itemize}

Based on the experimental observations presented above, we can summarize that our proposed methods have the following advantages: 1) Monotonic structures can better learn the correlation between $X^{s}$ and the hotel ranking score. 2) MSPL can better seek a balance between the ranking results based on user implicit feedback and merchant quality.

\subsection{Influence of Hyper-parameters $\lambda_{1}$ and $\lambda_{2}$}
\label{sec:hp}
The hyper-parameters $\lambda_1$ and $\lambda_2$ aim to balance CTRCVR and MCI ranking results. In order to explore the influence of these two hyper-parameters, we choose the CTCVR AUC and NDCG$@$20 as the measuring metrics. The range of $\lambda_1$ is in $\{0.1,0.5,1.0\}$ and the range of $\lambda_2$ is in $\{0.01,0.05,0.1,0.2\}$. The larger $\lambda_1$ will enlarge the influence of $\mathcal{L}^{CTRCVR}_{pairwise}$ while the larger $\lambda_2$ will enlarge the influence of $\mathcal{L}^{MCI|CTRCVR}_{pairwise}$. With the growth of $\lambda_2$, the result of CTCVR AUC gradually increases while NDCG@20 declines as shown in Figure \ref{fig:hp}. The influence of $\lambda_1$ is the opposite. From the online experiment, we consider that a 0.005 decline of CTCVR AUC is acceptable, so we set the lower bound of CTCVR AUC as 0.892 (The best CTCVR AUC is 0.897 as shown in Table \ref{table:auc}.). In order to choose the best hyper-parameters, we choose the points in the light green area (above the dashed line) and then we choose the point with the largest NDCG@20. Accordingly, we set $\lambda_1=1.0$ and $\lambda_2=0.1$ in Figure \ref{fig:hp}(c) as the appropriate hyper-parameters."

\subsection{Online A/B Test}
\begin{table}[htbp]
\caption{Online performance of the proposed MERIT model.}
\centering
\begin{tabular}{cccc}
\toprule \toprule
{}&{CVR }&{Sales Volume }&{MCI Score}\\
\midrule
{Lift}&{+1.50\%}&{+0.97\%}&{+3.02\%}\\ 
\bottomrule \bottomrule
\end{tabular}
\label{table:abtest}
\end{table}

We conduct an online A/B test in the hotel S\&R system on Fliggy within two weeks of July 2022. Table \ref{table:abtest} illustrates the online performance that our method is compared with the baseline method (DNN). It is apparent that our method is superior to the baseline method for both consumers and merchants. The CVR has increased by 1.50\%, and the Sales Volume has increased by 0.97\% (Given the substantial sales volume of the OTP, with hotel price in the hundreds of yuan, a 0.97\% increase in sales volume indicates a nearly one million yuan increase in sales volume.). The mean MCI score of displayed hotels has risen by 3.02\%, indicating that displayed merchant quality has been improved. Our proposed method has been deployed online and brought remarkable OTP profit growth.

\begin{figure}[htbp]
  \centering
  \includegraphics[width=\linewidth]{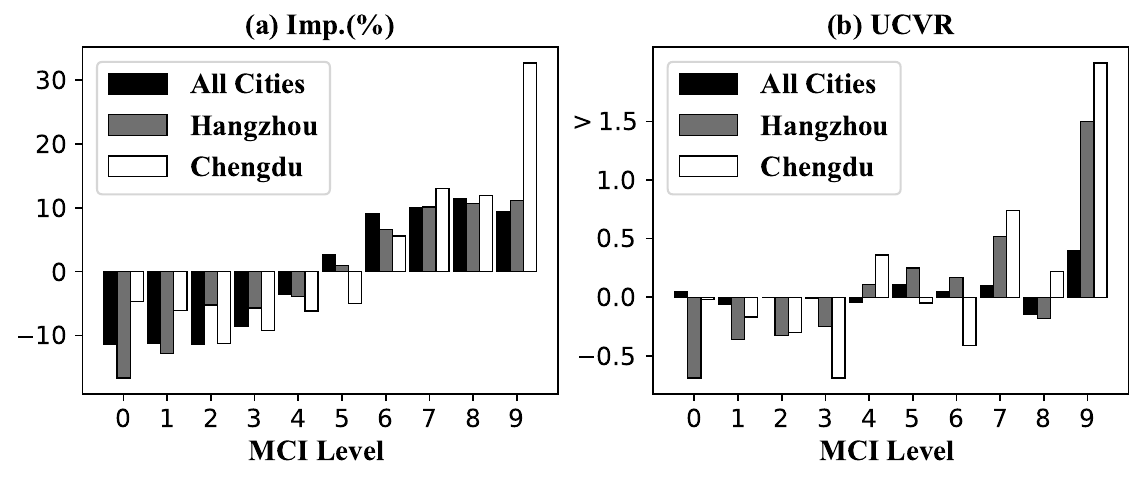}
  \caption{The online performance of MERIT model for hotels with different MCI levels among cities. The MCI level 0 denotes that there are no consumer ratings for hotels}
  \label{fig:abtest}
\end{figure}

We selected several major cities in China to compare the variations in the A/B test across different scenarios.  And we evaluated two significant online metrics, \textit{Imp.(\%)} and \textit{UCVR}, to assess the performance. \textit{Imp.(\%)} measures the lift ratio of the average displayed number for each hotel. \textit{UCVR} measures the proportion of users with orders among all users. Figure \ref{fig:abtest}(a) illustrates the \textit{Imp.(\%)} results for different MCI-level hotels in the online A/B test. With our proposed method, the results indicate that the MCI levels from 3.0 to 5.0 have an  improvement of impression in comparison to the base model. Hotels with an MCI level greater than 3.0 can, for example, achieve significant impression gains in Hangzhou, whereas hotels with an MCI level greater than 3.5 can achieve comparable impression gains in Chengdu. Additionally, Hangzhou's biggest growth is 19.31\% at MCI level 5.0, whereas Chengdu's largest growth is 32.66\% at MCI level 5.0. Similarly, in Figure \ref{fig:abtest}(b) we evaluate how \textit{UCVR} has changed for hotels with various MCI levels, and the results show that our proposed method has reallocated resources to hotels with high MCI levels. For example, hotels at the MCI level 5.0 have boosted their \textit{UCVR}s by 2.88\% in Hangzhou, 4.93\% in Chengdu, and 0.40\% in all cities. Therefore, we conclude that our proposed method has remarkable impression improvement on hotels with high MCI scores, which can address the former two challenges on OTPs: Matthew Effect in the consumer feedback-loop and the unclear relation between hotel quality and performance.

\section{Conclusion}
In this paper, we identify three main challenges in the scenario of the hotel S\&R system on Online Travel Platforms (OTPs): Matthew Effect in the consumer feedback-loop, unclear relation between hotel quality and performance, and conflicts between short-term and long-term revenue. A new Merchant Incentive Ranking Model representing the merchant-consumer relation for the hotel S\&R system, namely MERIT, is proposed to address these three challenges. We define factors of hotel quality, which represent the hotel quality of the entire consumer behavior cycle, and propose an MCI metric as an evaluation of hotel quality. Also, we introduce a monotonic structure into MERIT to provide clear relation between factors of hotel quality and ranking performance. Finally, we propose a novel Multi-objective Stratified Pairwise Loss to mitigate conflicts between platform revenue and consumer experience. Extensive experiment findings on the offline hotel S\&R dataset demonstrate the superiority of MERIT over existing state-of-the-art benchmarks in terms of NDCG, as well as the effectiveness of imposing monotonicity of merchant quality. We also conduct an online A/B test and obtain a 3.02\% improvement in the MCI score. MERIT has been deployed into Fliggy to serve tens of millions of consumers and hundreds of thousands of hotel merchants.





\bibliographystyle{ACM-Reference-Format}
\balance
\bibliography{sample-base}


\begin{thebibliography}{41}


\ifx \showCODEN    \undefined \def \showCODEN     #1{\unskip}     \fi
\ifx \showDOI      \undefined \def \showDOI       #1{#1}\fi
\ifx \showISBNx    \undefined \def \showISBNx     #1{\unskip}     \fi
\ifx \showISBNxiii \undefined \def \showISBNxiii  #1{\unskip}     \fi
\ifx \showISSN     \undefined \def \showISSN      #1{\unskip}     \fi
\ifx \showLCCN     \undefined \def \showLCCN      #1{\unskip}     \fi
\ifx \shownote     \undefined \def \shownote      #1{#1}          \fi
\ifx \showarticletitle \undefined \def \showarticletitle #1{#1}   \fi
\ifx \showURL      \undefined \def \showURL       {\relax}        \fi
\providecommand\bibfield[2]{#2}
\providecommand\bibinfo[2]{#2}
\providecommand\natexlab[1]{#1}
\providecommand\showeprint[2][]{arXiv:#2}

\bibitem[Archer and Wang(1993)]%
        {archer1993application}
\bibfield{author}{\bibinfo{person}{Norman~P Archer} {and}
  \bibinfo{person}{Shouhong Wang}.} \bibinfo{year}{1993}\natexlab{}.
\newblock \showarticletitle{Application of the back propagation neural network
  algorithm with monotonicity constraints for two-group classification
  problems}.
\newblock \bibinfo{journal}{\emph{Decision Sciences}} \bibinfo{volume}{24},
  \bibinfo{number}{1} (\bibinfo{year}{1993}), \bibinfo{pages}{60--75}.
\newblock


\bibitem[Barkan and Koenigstein(2016)]%
        {barkan2016item2vec}
\bibfield{author}{\bibinfo{person}{Oren Barkan} {and} \bibinfo{person}{Noam
  Koenigstein}.} \bibinfo{year}{2016}\natexlab{}.
\newblock \showarticletitle{Item2vec: neural item embedding for collaborative
  filtering}. In \bibinfo{booktitle}{\emph{2016 IEEE 26th International
  Workshop on Machine Learning for Signal Processing (MLSP)}}. IEEE,
  \bibinfo{pages}{1--6}.
\newblock


\bibitem[Bhattacharjee and Goel(2006)]%
        {b2006incentive}
\bibfield{author}{\bibinfo{person}{Rajat Bhattacharjee} {and}
  \bibinfo{person}{Ashish Goel}.} \bibinfo{year}{2006}\natexlab{}.
\newblock \showarticletitle{Incentive based ranking mechanisms}. In
  \bibinfo{booktitle}{\emph{First Workshop on the Economics of Networked
  Systems (Netecon’06)}}. \bibinfo{pages}{62--68}.
\newblock


\bibitem[Bhattacharjee et~al\mbox{.}(2009)]%
        {b2009incentive}
\bibfield{author}{\bibinfo{person}{Rajat Bhattacharjee},
  \bibinfo{person}{Ashish Goel}, {and} \bibinfo{person}{Konstantinos Kollias}.}
  \bibinfo{year}{2009}\natexlab{}.
\newblock \showarticletitle{An incentive-based architecture for social
  recommendations}. In \bibinfo{booktitle}{\emph{Proceedings of the third ACM
  conference on Recommender systems}}. \bibinfo{pages}{229--232}.
\newblock


\bibitem[Canini et~al\mbox{.}(2016)]%
        {milani2016fast}
\bibfield{author}{\bibinfo{person}{K. Canini}, \bibinfo{person}{A. Cotter},
  \bibinfo{person}{M.~R. Gupta}, \bibinfo{person}{M.~Milani Fard}, {and}
  \bibinfo{person}{J. Pfeifer}.} \bibinfo{year}{2016}\natexlab{}.
\newblock \showarticletitle{Fast and Flexible Monotonic Functions with
  Ensembles of Lattices}. In \bibinfo{booktitle}{\emph{Proceedings of the 30th
  International Conference on Neural Information Processing Systems}}
  \emph{(\bibinfo{series}{NIPS'16})}. \bibinfo{pages}{2927–2935}.
\newblock
\showISBNx{9781510838819}


\bibitem[Caruana(1997)]%
        {caruana1997hard_share}
\bibfield{author}{\bibinfo{person}{Rich Caruana}.}
  \bibinfo{year}{1997}\natexlab{}.
\newblock \showarticletitle{Multitask learning}.
\newblock \bibinfo{journal}{\emph{Machine learning}} \bibinfo{volume}{28},
  \bibinfo{number}{1} (\bibinfo{year}{1997}), \bibinfo{pages}{41--75}.
\newblock


\bibitem[Chen et~al\mbox{.}(2020)]%
        {chen2020bias}
\bibfield{author}{\bibinfo{person}{Jiawei Chen}, \bibinfo{person}{Hande Dong},
  \bibinfo{person}{Xiang Wang}, \bibinfo{person}{Fuli Feng},
  \bibinfo{person}{Meng Wang}, {and} \bibinfo{person}{Xiangnan He}.}
  \bibinfo{year}{2020}\natexlab{}.
\newblock \showarticletitle{Bias and debias in recommender system: A survey and
  future directions}.
\newblock \bibinfo{journal}{\emph{arXiv preprint arXiv:2010.03240}}
  (\bibinfo{year}{2020}).
\newblock


\bibitem[Cheng et~al\mbox{.}(2016)]%
        {cheng2016wideanddeep}
\bibfield{author}{\bibinfo{person}{Heng-Tze Cheng}, \bibinfo{person}{Levent
  Koc}, \bibinfo{person}{Jeremiah Harmsen}, \bibinfo{person}{Tal Shaked},
  \bibinfo{person}{Tushar Chandra}, \bibinfo{person}{Hrishi Aradhye},
  \bibinfo{person}{Glen Anderson}, \bibinfo{person}{Greg Corrado},
  \bibinfo{person}{Wei Chai}, \bibinfo{person}{Mustafa Ispir}, {et~al\mbox{.}}}
  \bibinfo{year}{2016}\natexlab{}.
\newblock \showarticletitle{Wide \& deep learning for recommender systems}. In
  \bibinfo{booktitle}{\emph{Proceedings of the 1st workshop on deep learning
  for recommender systems}}. \bibinfo{pages}{7--10}.
\newblock


\bibitem[Chung et~al\mbox{.}(2014)]%
        {chung2014GRU}
\bibfield{author}{\bibinfo{person}{Junyoung Chung}, \bibinfo{person}{Caglar
  Gulcehre}, \bibinfo{person}{KyungHyun Cho}, {and} \bibinfo{person}{Yoshua
  Bengio}.} \bibinfo{year}{2014}\natexlab{}.
\newblock \showarticletitle{Empirical evaluation of gated recurrent neural
  networks on sequence modeling}.
\newblock \bibinfo{journal}{\emph{arXiv preprint arXiv:1412.3555}}
  (\bibinfo{year}{2014}).
\newblock


\bibitem[Cole and Williamson(2019)]%
        {cole2019avoiding}
\bibfield{author}{\bibinfo{person}{Guy~W Cole} {and} \bibinfo{person}{Sinead~A
  Williamson}.} \bibinfo{year}{2019}\natexlab{}.
\newblock \showarticletitle{Avoiding resentment via monotonic fairness}.
\newblock \bibinfo{journal}{\emph{arXiv preprint arXiv:1909.01251}}
  (\bibinfo{year}{2019}).
\newblock


\bibitem[Daniels and Velikova(2010)]%
        {daniels2010minmax}
\bibfield{author}{\bibinfo{person}{Hennie Daniels} {and}
  \bibinfo{person}{Marina Velikova}.} \bibinfo{year}{2010}\natexlab{}.
\newblock \showarticletitle{Monotone and partially monotone neural networks}.
\newblock \bibinfo{journal}{\emph{IEEE Transactions on Neural Networks}}
  \bibinfo{volume}{21}, \bibinfo{number}{6} (\bibinfo{year}{2010}),
  \bibinfo{pages}{906--917}.
\newblock


\bibitem[Feelders(2000)]%
        {feelders2000prior}
\bibfield{author}{\bibinfo{person}{Ad~J Feelders}.}
  \bibinfo{year}{2000}\natexlab{}.
\newblock \showarticletitle{Prior knowledge in economic applications of data
  mining}. In \bibinfo{booktitle}{\emph{European Conference on Principles of
  Data Mining and Knowledge Discovery}}. Springer, \bibinfo{pages}{395--400}.
\newblock


\bibitem[Gardner and Dorling(1998)]%
        {gardner1998mlp}
\bibfield{author}{\bibinfo{person}{Matt~W Gardner} {and} \bibinfo{person}{SR
  Dorling}.} \bibinfo{year}{1998}\natexlab{}.
\newblock \showarticletitle{Artificial neural networks (the multilayer
  perceptron)—a review of applications in the atmospheric sciences}.
\newblock \bibinfo{journal}{\emph{Atmospheric environment}}
  \bibinfo{volume}{32}, \bibinfo{number}{14-15} (\bibinfo{year}{1998}),
  \bibinfo{pages}{2627--2636}.
\newblock


\bibitem[Guo et~al\mbox{.}(2017)]%
        {guo2017deepfm}
\bibfield{author}{\bibinfo{person}{Huifeng Guo}, \bibinfo{person}{Ruiming
  Tang}, \bibinfo{person}{Yunming Ye}, \bibinfo{person}{Zhenguo Li}, {and}
  \bibinfo{person}{Xiuqiang He}.} \bibinfo{year}{2017}\natexlab{}.
\newblock \showarticletitle{DeepFM: a factorization-machine based neural
  network for CTR prediction}.
\newblock \bibinfo{journal}{\emph{arXiv preprint arXiv:1703.04247}}
  (\bibinfo{year}{2017}).
\newblock


\bibitem[Gupta et~al\mbox{.}(2019)]%
        {gupta2019pointwiseloss}
\bibfield{author}{\bibinfo{person}{Akhil Gupta}, \bibinfo{person}{Naman
  Shukla}, \bibinfo{person}{Lavanya Marla}, \bibinfo{person}{Arinbj{\"o}rn
  Kolbeinsson}, {and} \bibinfo{person}{Kartik Yellepeddi}.}
  \bibinfo{year}{2019}\natexlab{}.
\newblock \showarticletitle{How to incorporate monotonicity in deep networks
  while preserving flexibility?}
\newblock \bibinfo{journal}{\emph{arXiv preprint arXiv:1909.10662}}
  (\bibinfo{year}{2019}).
\newblock


\bibitem[Haldar et~al\mbox{.}(2019)]%
        {haldar2019applying}
\bibfield{author}{\bibinfo{person}{Malay Haldar}, \bibinfo{person}{Mustafa
  Abdool}, \bibinfo{person}{Prashant Ramanathan}, \bibinfo{person}{Tao Xu},
  \bibinfo{person}{Shulin Yang}, \bibinfo{person}{Huizhong Duan},
  \bibinfo{person}{Qing Zhang}, \bibinfo{person}{Nick Barrow-Williams},
  \bibinfo{person}{Bradley~C Turnbull}, \bibinfo{person}{Brendan~M Collins},
  {et~al\mbox{.}}} \bibinfo{year}{2019}\natexlab{}.
\newblock \showarticletitle{Applying deep learning to airbnb search}. In
  \bibinfo{booktitle}{\emph{Proceedings of the 25th ACM SIGKDD International
  Conference on Knowledge Discovery \& Data Mining}}.
  \bibinfo{pages}{1927--1935}.
\newblock


\bibitem[Haldar et~al\mbox{.}(2020)]%
        {haldar2020improving}
\bibfield{author}{\bibinfo{person}{Malay Haldar}, \bibinfo{person}{Prashant
  Ramanathan}, \bibinfo{person}{Tyler Sax}, \bibinfo{person}{Mustafa Abdool},
  \bibinfo{person}{Lanbo Zhang}, \bibinfo{person}{Aamir Mansawala},
  \bibinfo{person}{Shulin Yang}, \bibinfo{person}{Bradley Turnbull}, {and}
  \bibinfo{person}{Junshuo Liao}.} \bibinfo{year}{2020}\natexlab{}.
\newblock \showarticletitle{Improving deep learning for airbnb search}. In
  \bibinfo{booktitle}{\emph{Proceedings of the 26th ACM SIGKDD International
  Conference on Knowledge Discovery \& Data Mining}}.
  \bibinfo{pages}{2822--2830}.
\newblock


\bibitem[He et~al\mbox{.}(2017)]%
        {he2017neural}
\bibfield{author}{\bibinfo{person}{Xiangnan He}, \bibinfo{person}{Lizi Liao},
  \bibinfo{person}{Hanwang Zhang}, \bibinfo{person}{Liqiang Nie},
  \bibinfo{person}{Xia Hu}, {and} \bibinfo{person}{Tat-Seng Chua}.}
  \bibinfo{year}{2017}\natexlab{}.
\newblock \showarticletitle{Neural collaborative filtering}. In
  \bibinfo{booktitle}{\emph{Proceedings of the 26th international conference on
  world wide web}}. \bibinfo{pages}{173--182}.
\newblock


\bibitem[Huang et~al\mbox{.}(2021)]%
        {dirn}
\bibfield{author}{\bibinfo{person}{Zai Huang}, \bibinfo{person}{Mingyuan Tao},
  {and} \bibinfo{person}{Bufeng Zhang}.} \bibinfo{year}{2021}\natexlab{}.
\newblock \showarticletitle{Deep Inclusion Relation-aware Network for User
  Response Prediction at Fliggy}. In \bibinfo{booktitle}{\emph{Proceedings of
  the 27th ACM SIGKDD Conference on Knowledge Discovery \& Data Mining}}.
  \bibinfo{pages}{3059--3067}.
\newblock


\bibitem[Jacobs et~al\mbox{.}(1991)]%
        {jacobs1991moe}
\bibfield{author}{\bibinfo{person}{Robert~A Jacobs}, \bibinfo{person}{Michael~I
  Jordan}, \bibinfo{person}{Steven~J Nowlan}, {and} \bibinfo{person}{Geoffrey~E
  Hinton}.} \bibinfo{year}{1991}\natexlab{}.
\newblock \showarticletitle{Adaptive mixtures of local experts}.
\newblock \bibinfo{journal}{\emph{Neural computation}} \bibinfo{volume}{3},
  \bibinfo{number}{1} (\bibinfo{year}{1991}), \bibinfo{pages}{79--87}.
\newblock


\bibitem[Liu et~al\mbox{.}(2020)]%
        {liu2020certified}
\bibfield{author}{\bibinfo{person}{Xingchao Liu}, \bibinfo{person}{Xing Han},
  \bibinfo{person}{Na Zhang}, {and} \bibinfo{person}{Qiang Liu}.}
  \bibinfo{year}{2020}\natexlab{}.
\newblock \showarticletitle{Certified monotonic neural networks}.
\newblock \bibinfo{journal}{\emph{Advances in Neural Information Processing
  Systems}}  \bibinfo{volume}{33} (\bibinfo{year}{2020}),
  \bibinfo{pages}{15427--15438}.
\newblock


\bibitem[Ma et~al\mbox{.}(2018b)]%
        {ma2018mmoe}
\bibfield{author}{\bibinfo{person}{Jiaqi Ma}, \bibinfo{person}{Zhe Zhao},
  \bibinfo{person}{Xinyang Yi}, \bibinfo{person}{Jilin Chen},
  \bibinfo{person}{Lichan Hong}, {and} \bibinfo{person}{Ed~H Chi}.}
  \bibinfo{year}{2018}\natexlab{b}.
\newblock \showarticletitle{Modeling task relationships in multi-task learning
  with multi-gate mixture-of-experts}. In \bibinfo{booktitle}{\emph{Proceedings
  of the 24th ACM SIGKDD international conference on knowledge discovery \&
  data mining}}. \bibinfo{pages}{1930--1939}.
\newblock


\bibitem[Ma et~al\mbox{.}(2018a)]%
        {ma2018entire}
\bibfield{author}{\bibinfo{person}{Xiao Ma}, \bibinfo{person}{Liqin Zhao},
  \bibinfo{person}{Guan Huang}, \bibinfo{person}{Zhi Wang},
  \bibinfo{person}{Zelin Hu}, \bibinfo{person}{Xiaoqiang Zhu}, {and}
  \bibinfo{person}{Kun Gai}.} \bibinfo{year}{2018}\natexlab{a}.
\newblock \showarticletitle{Entire space multi-task model: An effective
  approach for estimating post-click conversion rate}. In
  \bibinfo{booktitle}{\emph{The 41st International ACM SIGIR Conference on
  Research \& Development in Information Retrieval}}.
  \bibinfo{pages}{1137--1140}.
\newblock


\bibitem[Mikolov et~al\mbox{.}(2013)]%
        {mikolov2013word2vec}
\bibfield{author}{\bibinfo{person}{Tomas Mikolov}, \bibinfo{person}{Ilya
  Sutskever}, \bibinfo{person}{Kai Chen}, \bibinfo{person}{Greg~S Corrado},
  {and} \bibinfo{person}{Jeff Dean}.} \bibinfo{year}{2013}\natexlab{}.
\newblock \showarticletitle{Distributed representations of words and phrases
  and their compositionality}.
\newblock \bibinfo{journal}{\emph{Advances in neural information processing
  systems}}  \bibinfo{volume}{26} (\bibinfo{year}{2013}).
\newblock


\bibitem[Nguyen and Mart{\'\i}nez(2019)]%
        {nguyen2019mononet}
\bibfield{author}{\bibinfo{person}{An-phi Nguyen} {and}
  \bibinfo{person}{Mar{\'\i}a~Rodr{\'\i}guez Mart{\'\i}nez}.}
  \bibinfo{year}{2019}\natexlab{}.
\newblock \showarticletitle{Mononet: towards interpretable models by learning
  monotonic features}.
\newblock \bibinfo{journal}{\emph{arXiv preprint arXiv:1909.13611}}
  (\bibinfo{year}{2019}).
\newblock


\bibitem[Qu et~al\mbox{.}(2016)]%
        {PNN}
\bibfield{author}{\bibinfo{person}{Yanru Qu}, \bibinfo{person}{Han Cai},
  \bibinfo{person}{Kan Ren}, \bibinfo{person}{Weinan Zhang},
  \bibinfo{person}{Yong Yu}, \bibinfo{person}{Ying Wen}, {and}
  \bibinfo{person}{Jun Wang}.} \bibinfo{year}{2016}\natexlab{}.
\newblock \showarticletitle{Product-based neural networks for user response
  prediction}. In \bibinfo{booktitle}{\emph{2016 IEEE 16th International
  Conference on Data Mining (ICDM)}}. IEEE, \bibinfo{pages}{1149--1154}.
\newblock


\bibitem[Rendle(2010)]%
        {rendle2010factorization}
\bibfield{author}{\bibinfo{person}{Steffen Rendle}.}
  \bibinfo{year}{2010}\natexlab{}.
\newblock \showarticletitle{Factorization machines}. In
  \bibinfo{booktitle}{\emph{2010 IEEE International conference on data
  mining}}. IEEE, \bibinfo{pages}{995--1000}.
\newblock


\bibitem[Rendle et~al\mbox{.}(2012)]%
        {rendle2012bpr}
\bibfield{author}{\bibinfo{person}{Steffen Rendle}, \bibinfo{person}{Christoph
  Freudenthaler}, \bibinfo{person}{Zeno Gantner}, {and} \bibinfo{person}{Lars
  Schmidt-Thieme}.} \bibinfo{year}{2012}\natexlab{}.
\newblock \showarticletitle{BPR: Bayesian personalized ranking from implicit
  feedback}.
\newblock \bibinfo{journal}{\emph{arXiv preprint arXiv:1205.2618}}
  (\bibinfo{year}{2012}).
\newblock


\bibitem[Sculley et~al\mbox{.}(2015)]%
        {nips2015feedbackloop}
\bibfield{author}{\bibinfo{person}{D. Sculley}, \bibinfo{person}{Gary Holt},
  \bibinfo{person}{Daniel Golovin}, \bibinfo{person}{Eugene Davydov},
  \bibinfo{person}{Todd Phillips}, \bibinfo{person}{Dietmar Ebner},
  \bibinfo{person}{Vinay Chaudhary}, \bibinfo{person}{Michael Young},
  \bibinfo{person}{Jean{-}Fran{\c{c}}ois Crespo}, {and} \bibinfo{person}{Dan
  Dennison}.} \bibinfo{year}{2015}\natexlab{}.
\newblock \showarticletitle{Hidden Technical Debt in Machine Learning Systems}.
  In \bibinfo{booktitle}{\emph{Advances in Neural Information Processing
  Systems 28: Annual Conference on Neural Information Processing Systems}}.
  \bibinfo{pages}{2503--2511}.
\newblock


\bibitem[Simchi-Levi et~al\mbox{.}(2022)]%
        {levi2022nonparametric}
\bibfield{author}{\bibinfo{person}{David Simchi-Levi}, \bibinfo{person}{Rui
  Sun}, \bibinfo{person}{Michelle Wu}, {and} \bibinfo{person}{Ruihao Zhu}.}
  \bibinfo{year}{2022}\natexlab{}.
\newblock \showarticletitle{Calibrating Sales Forecast in a Pandemic Using
  Competitive Online Non-Parametric Regression}. In
  \bibinfo{booktitle}{\emph{SSRN 3670264}}.
\newblock


\bibitem[Tang et~al\mbox{.}(2020a)]%
        {tang2020ple}
\bibfield{author}{\bibinfo{person}{Hongyan Tang}, \bibinfo{person}{Junning
  Liu}, \bibinfo{person}{Ming Zhao}, {and} \bibinfo{person}{Xudong Gong}.}
  \bibinfo{year}{2020}\natexlab{a}.
\newblock \showarticletitle{Progressive layered extraction (ple): A novel
  multi-task learning (mtl) model for personalized recommendations}. In
  \bibinfo{booktitle}{\emph{Fourteenth ACM Conference on Recommender Systems}}.
  \bibinfo{pages}{269--278}.
\newblock


\bibitem[Tang et~al\mbox{.}(2020b)]%
        {ple}
\bibfield{author}{\bibinfo{person}{Hongyan Tang}, \bibinfo{person}{Junning
  Liu}, \bibinfo{person}{Ming Zhao}, {and} \bibinfo{person}{Xudong Gong}.}
  \bibinfo{year}{2020}\natexlab{b}.
\newblock \showarticletitle{Progressive layered extraction (ple): A novel
  multi-task learning (mtl) model for personalized recommendations}. In
  \bibinfo{booktitle}{\emph{Fourteenth ACM Conference on Recommender Systems}}.
  \bibinfo{pages}{269--278}.
\newblock


\bibitem[Wang et~al\mbox{.}(2017)]%
        {wang2017dcn}
\bibfield{author}{\bibinfo{person}{Ruoxi Wang}, \bibinfo{person}{Bin Fu},
  \bibinfo{person}{Gang Fu}, {and} \bibinfo{person}{Mingliang Wang}.}
  \bibinfo{year}{2017}\natexlab{}.
\newblock \showarticletitle{Deep \& cross network for ad click predictions}.
\newblock In \bibinfo{booktitle}{\emph{Proceedings of the ADKDD'17}}.
  \bibinfo{pages}{1--7}.
\newblock


\bibitem[Wen et~al\mbox{.}(2020)]%
        {wen2020entire}
\bibfield{author}{\bibinfo{person}{Hong Wen}, \bibinfo{person}{Jing Zhang},
  \bibinfo{person}{Yuan Wang}, \bibinfo{person}{Fuyu Lv},
  \bibinfo{person}{Wentian Bao}, \bibinfo{person}{Quan Lin}, {and}
  \bibinfo{person}{Keping Yang}.} \bibinfo{year}{2020}\natexlab{}.
\newblock \showarticletitle{Entire space multi-task modeling via post-click
  behavior decomposition for conversion rate prediction}. In
  \bibinfo{booktitle}{\emph{Proceedings of the 43rd International ACM SIGIR
  conference on research and development in Information Retrieval}}.
  \bibinfo{pages}{2377--2386}.
\newblock


\bibitem[Wu et~al\mbox{.}(2022b)]%
        {wu2022cheaper}
\bibfield{author}{\bibinfo{person}{Han Wu}, \bibinfo{person}{Hongzhe Zhang},
  \bibinfo{person}{Liangyue Li}, \bibinfo{person}{Zulong Chen},
  \bibinfo{person}{Fanwei Zhu}, {and} \bibinfo{person}{Xiao Fang}.}
  \bibinfo{year}{2022}\natexlab{b}.
\newblock \showarticletitle{Cheaper Is Better: Exploring Price Competitiveness
  for Online Purchase Prediction}. In \bibinfo{booktitle}{\emph{2022 IEEE 38th
  International Conference on Data Engineering (ICDE)}}. IEEE,
  \bibinfo{pages}{3399--3412}.
\newblock


\bibitem[Wu et~al\mbox{.}(2022a)]%
        {wu2022agde}
\bibfield{author}{\bibinfo{person}{Kailun Wu}, \bibinfo{person}{Weijie Bian},
  \bibinfo{person}{Zhangming Chan}, \bibinfo{person}{Lejian Ren},
  \bibinfo{person}{Shiming Xiang}, \bibinfo{person}{Shu-Guang Han},
  \bibinfo{person}{Hongbo Deng}, {and} \bibinfo{person}{Bo Zheng}.}
  \bibinfo{year}{2022}\natexlab{a}.
\newblock \showarticletitle{Adversarial Gradient Driven Exploration for Deep
  Click-Through Rate Prediction} \emph{(\bibinfo{series}{KDD '22})}.
  \bibinfo{pages}{2050–2058}.
\newblock


\bibitem[Xu et~al\mbox{.}(2022)]%
        {g2net}
\bibfield{author}{\bibinfo{person}{Jia Xu}, \bibinfo{person}{Fei Xiong},
  \bibinfo{person}{Zulong Chen}, \bibinfo{person}{Mingyuan Tao},
  \bibinfo{person}{Liangyue Li}, {and} \bibinfo{person}{Quan Lu}.}
  \bibinfo{year}{2022}\natexlab{}.
\newblock \showarticletitle{G2NET: A General Geography-Aware Representation
  Network for Hotel Search Ranking}. In \bibinfo{booktitle}{\emph{Proceedings
  of the 28th ACM SIGKDD Conference on Knowledge Discovery and Data Mining}}.
  \bibinfo{pages}{4237--4247}.
\newblock


\bibitem[You et~al\mbox{.}(2017)]%
        {you2017dln}
\bibfield{author}{\bibinfo{person}{Seungil You}, \bibinfo{person}{David Ding},
  \bibinfo{person}{Kevin Canini}, \bibinfo{person}{Jan Pfeifer}, {and}
  \bibinfo{person}{Maya~R. Gupta}.} \bibinfo{year}{2017}\natexlab{}.
\newblock \showarticletitle{Deep Lattice Networks and Partial Monotonic
  Functions}. In \bibinfo{booktitle}{\emph{Proceedings of the 31st
  International Conference on Neural Information Processing Systems}}
  \emph{(\bibinfo{series}{NIPS'17})}. \bibinfo{pages}{2985–2993}.
\newblock
\showISBNx{9781510860964}


\bibitem[Zhang et~al\mbox{.}(2022)]%
        {zhang2022price}
\bibfield{author}{\bibinfo{person}{Xiaokun Zhang}, \bibinfo{person}{Bo Xu},
  \bibinfo{person}{Liang Yang}, \bibinfo{person}{Chenliang Li},
  \bibinfo{person}{Fenglong Ma}, \bibinfo{person}{Haifeng Liu}, {and}
  \bibinfo{person}{Hongfei Lin}.} \bibinfo{year}{2022}\natexlab{}.
\newblock \showarticletitle{Price DOES Matter! Modeling Price and Interest
  Preferences in Session-based Recommendation}.
\newblock \bibinfo{journal}{\emph{arXiv preprint arXiv:2205.04181}}
  (\bibinfo{year}{2022}).
\newblock


\bibitem[Zhou et~al\mbox{.}(2019)]%
        {dien}
\bibfield{author}{\bibinfo{person}{Guorui Zhou}, \bibinfo{person}{Na Mou},
  \bibinfo{person}{Ying Fan}, \bibinfo{person}{Qi Pi}, \bibinfo{person}{Weijie
  Bian}, \bibinfo{person}{Chang Zhou}, \bibinfo{person}{Xiaoqiang Zhu}, {and}
  \bibinfo{person}{Kun Gai}.} \bibinfo{year}{2019}\natexlab{}.
\newblock \showarticletitle{Deep interest evolution network for click-through
  rate prediction}. In \bibinfo{booktitle}{\emph{Proceedings of the AAAI
  conference on artificial intelligence}}, Vol.~\bibinfo{volume}{33}.
  \bibinfo{pages}{5941--5948}.
\newblock


\bibitem[Zhou et~al\mbox{.}(2018)]%
        {din}
\bibfield{author}{\bibinfo{person}{Guorui Zhou}, \bibinfo{person}{Xiaoqiang
  Zhu}, \bibinfo{person}{Chenru Song}, \bibinfo{person}{Ying Fan},
  \bibinfo{person}{Han Zhu}, \bibinfo{person}{Xiao Ma},
  \bibinfo{person}{Yanghui Yan}, \bibinfo{person}{Junqi Jin},
  \bibinfo{person}{Han Li}, {and} \bibinfo{person}{Kun Gai}.}
  \bibinfo{year}{2018}\natexlab{}.
\newblock \showarticletitle{Deep interest network for click-through rate
  prediction}. In \bibinfo{booktitle}{\emph{Proceedings of the 24th ACM SIGKDD
  international conference on knowledge discovery \& data mining}}.
  \bibinfo{pages}{1059--1068}.
\newblock


\end{thebibliography}

\appendix

\end{document}